\documentclass[aps,prb,twocolumn,groupedaddress,showpacs]{revtex4} 

\usepackage{amssymb}
\usepackage{graphicx}
\usepackage{amsmath}
\usepackage{dsfont}
\usepackage{bbm}
\newcommand{\calt}{{\cal T}}
\newcommand{\cG}{\check{G}}
\newcommand{\cK}{\check{K}}
\newcommand{\cSigma}{\check{\Sigma}}
\newcommand{\cLambda}{\check{\Lambda}}
\newcommand{\cGamma}{\check{\Gamma}}
\newcommand{\cXi}{\check{\Xi}}
\newcommand{\trace}[1]{\text{Tr}_{\text{#1}}}

\begin{document}

\title{Current correlations in the interacting Cooper-pair beam-splitter}

\author{
	J.~Rech,
	D.~Chevallier,
	T.~Jonckheere, and
	T.~Martin
}

\affiliation{
	Centre de Physique Th\'eorique,  CNRS UMR 7332, Aix-Marseille Universit\'e, Case 907, 13288 Marseille, France}

\date{\today}

\begin{abstract}
We propose an approach allowing the computation of currents and their correlations in interacting multi-terminal mesoscopic systems involving quantum dots coupled to normal and/or superconducting leads. The formalism relies on the expression of branching currents and noise crossed correlations in terms of one- and two-particle Green's functions for the dots electrons, which are then evaluated self-consistently within a conserving approximation.
We then apply this to the Cooper-pair beam-splitter setup recently proposed~\cite{hofstetter,herrmann}, which we model as a double quantum dot with weak interactions, connected to a superconducting lead and two normal ones. Our method not only enables us to take into account a local repulsive interaction on the dots, but also to study its competition with the direct tunneling between dots.
Our results suggest that even a weak Coulomb repulsion tends to favor positive current cross-correlations in the antisymmetric regime (where the dots have opposite energies with respect to the superconducting chemical potential).
\end{abstract}

\pacs{
	73.23.-b,   %Electronic transport in mesoscopic systems
	73.63.Kv    %Electronic transport in nanoscale materials and structures - Quantum dots
	74.45.+c, 	%Proximity effects; Andreev reflection; SN and SNS junctions
	72.70.+m    %Noise process and phenomena 
	71.10.-w    %Theories and models of many-electron systems
}

\maketitle

\section{Introduction} \label{sec:introduction}

With the development of nanofabrication techniques, more experiments in mesoscopic physics draw inspiration from quantum optics. Among them, there is a particular interest in generating entangled electronic states. Beyond the exploration of non-local quantum effects through the test of Bell's inequalities in a solid-state device, there are potential applications in quantum teleportation and information processing, where such sources of entangled pairs could be integrated with existing technology and infrastructure.

Because of the spin singlet character of Cooper pairs, superconductors constitute a natural source of spin-entangled Einstein-Podolsky-Rosen (EPR) pairs. The difficulty then lies in spatially separating the constituents of these pairs coherently. The commonly proposed setup to accomplish such a task is composed of a superconductor connected to a fork made out of two normal metal leads~\cite{anatram_datta,lesovik_martin_blatter,bouchiat,torres-martin}. The mechanism for such a Cooper-pair beam splitter then relies on crossed Andreev reflexion (CAR), a process in which the two electrons of a Cooper pair are sent into different normal electrodes~\cite{torres-martin,deutscher_feinberg,falci_feinberg_hekking} (as opposed to the conventional direct Andreev reflexion which is local). 

Previous attempts on metallic structures revealed the difficulty of distinguishing CAR form other contributions as there are various parasitic processes~\cite{beckman_prl,klapwijk,chandrasekhar}. Several theoretical works have suggested different directions in order to provide preferential enhancement of CAR, from simple spin and energy filtering to the effect of strong electron interactions, susceptible to favor single-particle over pair tunneling~\cite{sukhorukov_recher_loss,borlin,yeyati_bergeret,zaikin_golubev,eldridge}.

An important step in the observation of Cooper pair splitting came from recent experimental works in tunable double quantum dot devices based on carbon nanotubes~\cite{herrmann} and InAs nanowires~\cite{hofstetter}. These systems showed great promise as they not only display local Coulomb repulsion on the dots, but also allow for the exploration of different energy configurations of the dots, two properties that should promote CAR over other processes. 

Nevertheless, further efforts are needed both experimentally and theoretically. Future works should focus on time-resolved correlation measurements of the currents in the normal leads, particularly relevant to characterize non-local effects. However, to this date, they remain to be measured experimentally, and still represent an important challenge for theory, as there are only few examples of interacting mesoscopic devices driven out-of equilibrium, where current correlations have been obtained~\cite{hershfield,lopez,thielmann,fujii,simon}, none of which involving hybrid systems.

In this work, we propose a general approach for computing currents and current-current correlations in an interacting multi-terminal mesoscopic device involving quantum dots coupled to multiple normal and superconducting leads. While it is known\cite{non-int} that these quantities can be expressed in terms of the single and two-particle Green's functions of the dots electrons, here we show that those Green's functions can be computed self-consistently through a $\Phi$-derivable approximation, thereby ensuring that all conservation laws are respected. The method relies on a perturbation theory as formulated by Kadanoff and Baym~\cite{baym-kadanoff}, extended to the usual Keldysh contour to allow for a non-equilibrium situation, and will be referred to as the Kadanoff-Baym-Keldysh (KBK) approach.

When applying this KBK formalism to the double-dot Cooper-pair beam splitter setup, we show that even weak interactions have an important effect on both current and cross-correlations in the CAR-dominated regime, and that there exists a competition between Coulomb interaction and direct inter-dot tunneling. 

The outline of the paper is as follows. In section \ref{sec:model}, we discuss the general model for a set of quantum dots connected to multiple leads, and derive the corresponding expression for currents and current-current correlations. We then show in section \ref{sec:derivation} how the KBK approach allows us to compute the single- and two-particle Green's function. In section \ref{sec:application}, we apply this formalism to the case of the Cooper-pair beam-splitter. We conclude in section \ref{sec:conclusion}.

\section{Model} \label{sec:model}

Our starting point is a set of quantum dots (labeled $\alpha$), with energies $\epsilon_\alpha$, tunnel-coupled to multiple leads (labeled $j$) with tunneling amplitudes $t_{\alpha j}$. The leads have identical chemical potentials $\mu_j = \mu$ and are characterized by their voltage bias $V_j$ and superconducting order parameter $\Delta_j$ (normal leads correspond to $\Delta_j=0$). The Hamiltonian for such a system is of the form
\begin{equation}
H= \sum_{\alpha} H_\alpha + \sum_{j} H_j + \sum_{j,\alpha} H_{j\alpha}  .
\label{eq-Hamilt}
\end{equation}

In terms of Nambu spinors $\hat{d}_\alpha^\dagger = ( d_{\alpha \uparrow}^\dagger ~ d_{\alpha \downarrow} )$, the dots Hamiltonians read
\begin{equation}
H_\alpha = \epsilon_\alpha \hat{d}_\alpha^\dagger \sigma_z \hat{d}_\alpha + \sum_{\beta \neq \alpha} \left( \frac{t_{\alpha \beta}}{2} \hat{d}_\alpha^\dagger \sigma_z \hat{d}_\beta + \text{h.c.} \right) + U_\alpha n_{\alpha \uparrow} n_{\alpha \downarrow} ,\label{eq-HamiltDot}
\end{equation}
where we also introduced the inter-dot tunneling amplitude $t_{\alpha\beta}$ and the local Coulomb repulsion $U_\alpha$.

Similarly, the leads Hamiltonians are given by
\begin{equation}
H_j = \sum_k \hat{\Psi}_{jk}^\dagger \left( \xi_k \sigma_z + \Delta_j \sigma_x\right) \hat{\Psi}_{jk}
\label{eq-HamiltLead}
\end{equation}
where $\xi_k = \frac{k^2}{2m} - \mu$ and we used the Nambu spinors $\hat{\Psi}_{jk}^\dagger = ( \Psi_{jk\uparrow}^\dagger ~ \Psi_{j -k \downarrow} )$.  Here and throughout the remainder of the text, we work with physical dimensions corresponding to $\hbar=1$.

We chose to transfer the voltage dependence onto the tunneling term using the Peierls substitution: $\calt_{j \alpha} (t) = t_{j\alpha} \sigma_z e^{i \sigma_z V_j t} $. The tunneling Hamiltonian is thus given by
\begin{equation}
H_{j\alpha} = \sum_k \left( \hat{\Psi}_{jk}^\dagger \calt_{j \alpha} (t) \hat{d}_\alpha + \text{h.c.} \right) .
\label{eq-HamiltTunnel}
\end{equation}
 
Since the total Hamiltonian is quadratic in the leads electrons operators, we can integrate out these degrees of freedom. The effect of the leads is then captured by a tunneling self-energy $\cSigma_T (t_1,t_2) = \sum_j \cSigma_j (t_1,t_2)$,
\begin{align}
\cSigma_{j,\mu_1 \mu_2} (t_1,t_2) = \left[\tilde{\calt}_{j \alpha_1}^\dagger (t_1) ~\tilde{g}_j (t_1,t_2)~ \tilde{\calt}_{j \alpha_2} (t_2) \right]_{\sigma_1 \sigma_2}^{\eta_1 \eta_2} ,
\label{eq-TunnSelfE}
\end{align}
which depends on the tunneling amplitude written in Nambu-Keldysh space $\tilde{\calt}_{j \alpha} (t)= \tau_z \otimes \calt_{j \alpha} (t)$ as well as the standard non-interacting Green's function for lead electrons, given in Nambu-Keldysh space by
\begin{equation}
\tilde{g}_{j,\sigma_1 \sigma_2}^{\eta_1 \eta_2} (t_1,t_2) = -i \sum_k \langle T_K \hat{\Psi}_{jk\sigma_1} (t_1^{\eta_1}) \hat{\Psi}_{jk\sigma_2}^\dagger (t_2^{\eta_2}) \rangle . 
\end{equation} 
For convenience, we also introduced the more compact notation $\mu_i \equiv \left\{ \sigma_i, \alpha_i, \eta_i \right\}$, condensing the Nambu-dot-Keldysh components into a single index.

This averaging over the leads degrees of freedom allows us to express the relevant physical quantities in terms of the tunneling self-energy as well as the dots electrons single- and two-particle Green's functions. These are defined in Nambu-dot-Keldysh (NDK) space respectively as
\begin{align}
&\cG_{\mu_1 \mu_2} (t_1,t_2) = -i \langle T_K \hat{d}_{\alpha_1 \sigma_1} (t_1^{\eta_1}) \hat{d}_{\alpha_2 \sigma_2}^\dagger (t_2^{\eta_2}) \rangle ,
\end{align}
and
\begin{align}
&\cK_{\mu_1 \mu_2 \mu_3 \mu_4} (t_1,t_2,t_3,t_4) = \nonumber \\
&-\langle T_K \hat{d}_{\alpha_1 \sigma_1} (t_1^{\eta_1}) \hat{d}_{\alpha_2 \sigma_2} (t_2^{\eta_2}) \hat{d}_{\alpha_3 \sigma_3}^\dagger (t_3^{\eta_3}) \hat{d}_{\alpha_4 \sigma_4}^\dagger (t_4^{\eta_4})\rangle ,
\end{align}
where $T_K$ corresponds to the time-ordering operator along the Keldysh contour.

In particular, the average current from the dot $\alpha$ into the lead $j$ is readily obtained using standard methods~\cite{non-int} as a function of the single-particle Green's function $\cG$
\begin{align}
\langle I_{j\alpha} (t) \rangle =& \frac{e}{2} \trace{NDK} \bigg[ (\tau_z \otimes {\cal I}_\alpha \otimes \sigma_z) \nonumber \\
& \times \left. \int dt' \left( \cG (t,t') \cSigma_j (t',t) - \cSigma_j (t,t') \cG (t',t) \right) \right] ,
\label{eq-Current}
\end{align}
where the matrix ${\cal I}_\alpha$ is defined in dot space, with elements $\left[{\cal I}_\alpha\right]_{\alpha_1 \alpha_2} = \delta_{\alpha \alpha_1} \delta_{\alpha \alpha_2}$, and $\trace{NDK}$ corresponds to the trace in the full Nambu-dot-Keldysh space.

Similarly the current correlations can be expressed in terms of the two-particle Green's function $\cK$ as
\begin{multline}
\langle I_{i\alpha}^\eta (t) I_{j\beta}^{\eta'} (t') \rangle = e^2 \sum_{\sigma \sigma'} \sigma_z^{\sigma \sigma} \sigma_z^{\sigma' \sigma'} \int dt_1 dt_2  \sum_{\mu_1 \mu_2} \\
 \times \Big[ \cSigma_{i,\nu \mu_1} (t,t_1) \cSigma_{j,\nu' \mu_2} (t',t_2) \cK_{\mu_1 \mu_2 \nu \nu'} (t_1,t_2,t,t')  \\
-\cSigma_{i,\nu \mu_1} (t,t_1) \cSigma_{j,\mu_2 \nu'} (t_2,t') \cK_{\mu_1 \nu' \nu \mu_2} (t_1,t',t,t_2)  \\
-\cSigma_{i,\mu_1 \nu} (t_1,t) \cSigma_{j,\nu' \mu_2} (t',t_2) \cK_{\nu \mu_2 \mu_1 \nu'} (t,t_2,t_1,t')  \\
+\cSigma_{i,\mu_1 \nu} (t_1,t) \cSigma_{j,\mu_2 \nu'} (t_2,t') \cK_{\nu \nu' \mu_1 \mu_2} (t,t',t_1,t_2) \Big] ,
\label{eq-CurrCorr}
\end{multline}
with the labels $\nu \equiv \left\{ \sigma,\alpha,\eta\right\}$ and $\nu' \equiv \left\{ \sigma',\beta,\eta' \right\}$.

\section{Kadanoff-Baym-Keldysh formalism} \label{sec:derivation}

\begin{figure}[tb]
\centering
 \resizebox{.48\textwidth}{!}{
\begin{tabular}{r c} 
(a)& \includegraphics[width=8cm]{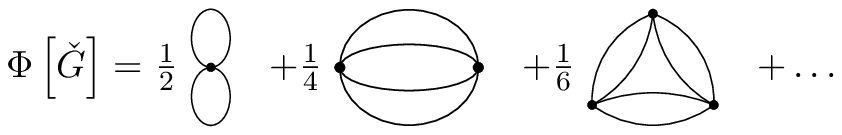} \\
(b)& \includegraphics[width=8cm]{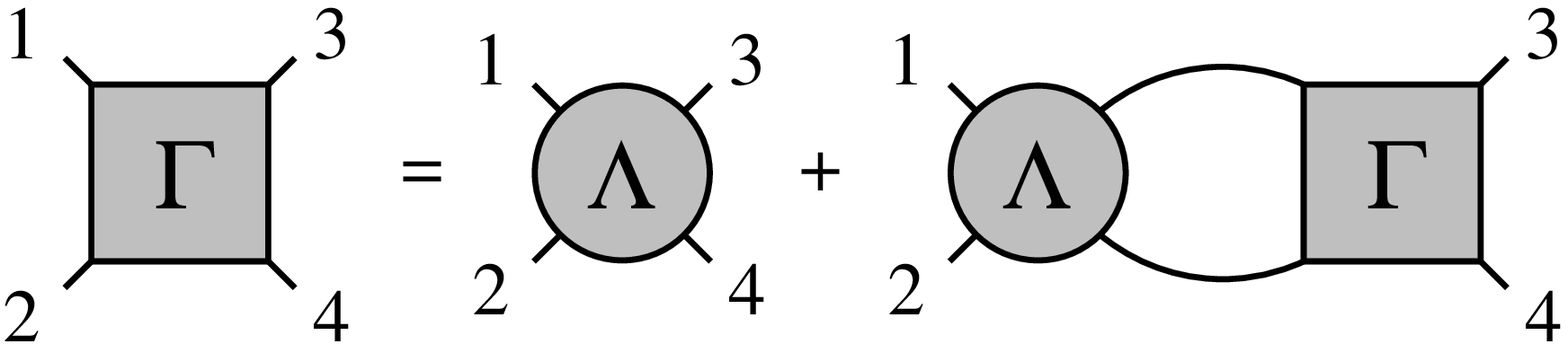} \\
(c)& \includegraphics[width=8cm]{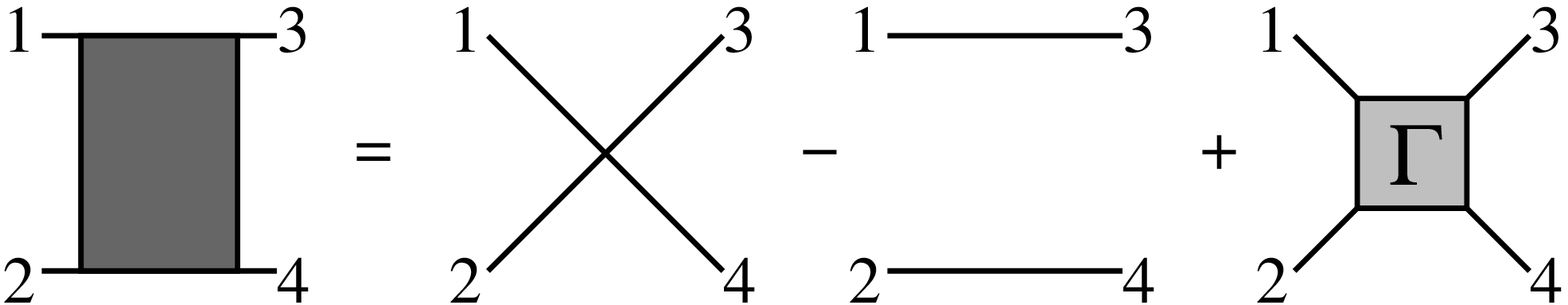}
 \end{tabular}}
\caption{
Diagrammatic representation of (a) the contributions to the Luttinger-Ward functional, (b) the Bethe-Salpeter equation in the horizontal particle-hole channel, and (c) the two-particle Green's function in terms of the single-particle one and the vertex function. The full lines correspond to full single-particle Green's functions for the dot electrons, the disks represent bare vertices. The large circles and squares correspond to the irreducible and full vertices respectively, while the shaded block stands for the two-particle Green's function $\cK$ defined in the text.}
\label{fig-diagrams}
\end{figure}

In order to proceed further, we need to evaluate the fully interacting dots electrons Green's functions $\cG$ and $\cK$. This is achieved through a perturbative expansion in the Coulomb interaction $U_\alpha$, within a so-called $\Phi$-derivable approximation\cite{baym-kadanoff}. Such a conserving approximation relies on the truncation of the Luttinger-Ward \cite{luttinger-ward} functional $\Phi \left[\cG\right]$. The latter is a function of the fully interacting single-particle Green's function $\cG$, and corresponds to the sum of all closed-loop two-particle irreducible diagrams. Unfortunately, there is no simple closed form of the Luttinger-Ward functional, which is usually represented in a more convenient diagrammatic form (see Fig.~\ref{fig-diagrams}a).

The first striking feature of the Luttinger-Ward functional is that it serves as a generating functional for the interacting self-energy. Indeed, taking the functional derivative of $\Phi \left[\cG\right]$ with respect to the one-particle Green's function $\cG$ leads to
\begin{equation}
\cSigma_{C,\mu_1 \mu_2} (t_1,t_2) = \frac{\delta \Phi \left[\cG\right]}{\delta \cG_{\mu_2 \mu_1} (t_2,t_1)} .
\end{equation}
Note that the resulting expression for the self-energy is itself a function of the full propagator, so that $\cG$ has to be derived self-consistently. This is achieved by solving the Dyson equation
\begin{align}
\cG (t_1,t_2) =& \cG^0 (t_1,t_2) + \int dt_3 dt_4 \cG^0 (t_1,t_3) \nonumber \\
&\times \left[ \cSigma_{C} (t_3,t_4) + \cSigma_T (t_3,t_4) \right] \cG (t_4,t_2),
\label{eq-Dyson}
\end{align}
where $\cG^0$ is the dots electrons Green's function in absence of both interactions and tunneling to the leads.

Further differentiating the Luttinger-Ward functional gives access to more involved many-body objects. In particular, one can derive an expression for the irreducible vertex from a second order functional differentiation. In the horizontal particle-hole\footnote{Note that here "particle" is used in the Nambu sense and could correspond to either electron or hole depending on the spin projection.} channel, it reads
\begin{align}
\cLambda_{\mu_1\mu_2\mu_3\mu_4} (t_1,t_2,t_3,t_4) = - \frac{\delta^2 \Phi \left[\cG\right]}{\delta \cG_{\mu_2 \mu_1} (t_2,t_1) \delta \cG_{\mu_3 \mu_4} (t_3,t_4)}.
\label{eq-irred}
\end{align}
This in turn allows us to write the full vertex function $\cGamma$ by self-consistently solving the Bethe-Salpeter equation in the corresponding irreducibility channel (see Fig.~\ref{fig-diagrams}b)
\begin{widetext}
\begin{align}
\cGamma_{\mu_1\mu_2\mu_3\mu_4} (t_1,t_2,t_3,t_4) &= \cLambda_{\mu_1\mu_2\mu_3\mu_4} (t_1,t_2,t_3,t_4) \nonumber \\
&+ \int dt_5 \ldots dt_8 \sum_{\mu_5 \ldots \mu_8} \cLambda_{\mu_1\mu_2\mu_5\mu_6} (t_1,t_2,t_5,t_6) \cG_{\mu_5 \mu_7} (t_5,t_7) \cG_{\mu_8 \mu_6} (t_8,t_6) \cGamma_{\mu_7\mu_8\mu_3\mu_4} (t_7,t_8,t_3,t_4) .
\label{eq-BetheSalpeter}
\end{align}
The two-particle dots electrons Green's function $\cK$ is then readily obtained from
\begin{align}
\cK_{\mu_1 \mu_2 \mu_3 \mu_4} (t_1,t_2,t_3,t_4) &= \cG_{\mu_1 \mu_4} (t_1,t_4) \cG_{\mu_2 \mu_3} (t_2,t_3) 
- \cG_{\mu_1 \mu_3} (t_1,t_3) \cG_{\mu_2 \mu_4} (t_2,t_4) \nonumber \\
&+ \sum_{\mu_5 \ldots \mu_8} \int dt_5 \ldots dt_8  \cG_{\mu_1 \mu_5} (t_1,t_5) \cG_{\mu_2 \mu_8} (t_2,t_8) \cGamma_{\mu_5\mu_6\mu_7\mu_8} (t_5,t_6,t_7,t_8) 
 \cG_{\mu_7 \mu_3} (t_7,t_3) \cG_{\mu_6 \mu_4} (t_6,t_4).
\label{eq-GammatoK}
\end{align}
\end{widetext}
It follows that from a given truncation of $\Phi \left[ \cG \right]$, one can determine both the full single- and two-particle Green's function, respectively via the derivative of the Luttinger-Ward functional and the self-consistent solution of the Dyson equation, then through the second derivative of $\Phi \left[ \cG \right]$, after solving the Bethe-Salpeter equation. Note that although both Dyson and Bethe-Salpeter equations have to be solved self-consistently, the proposed scheme is only one-particle self-consistent since the Luttinger-Ward functional is a function of $\cG$ only.

\section{Application to the Cooper-pair beam-splitter} \label{sec:application}

\begin{figure}[tb]
\centering
 \resizebox{.48\textwidth}{!}{
\begin{tabular}{c c c c} 
(a)& \includegraphics[width=4.8cm]{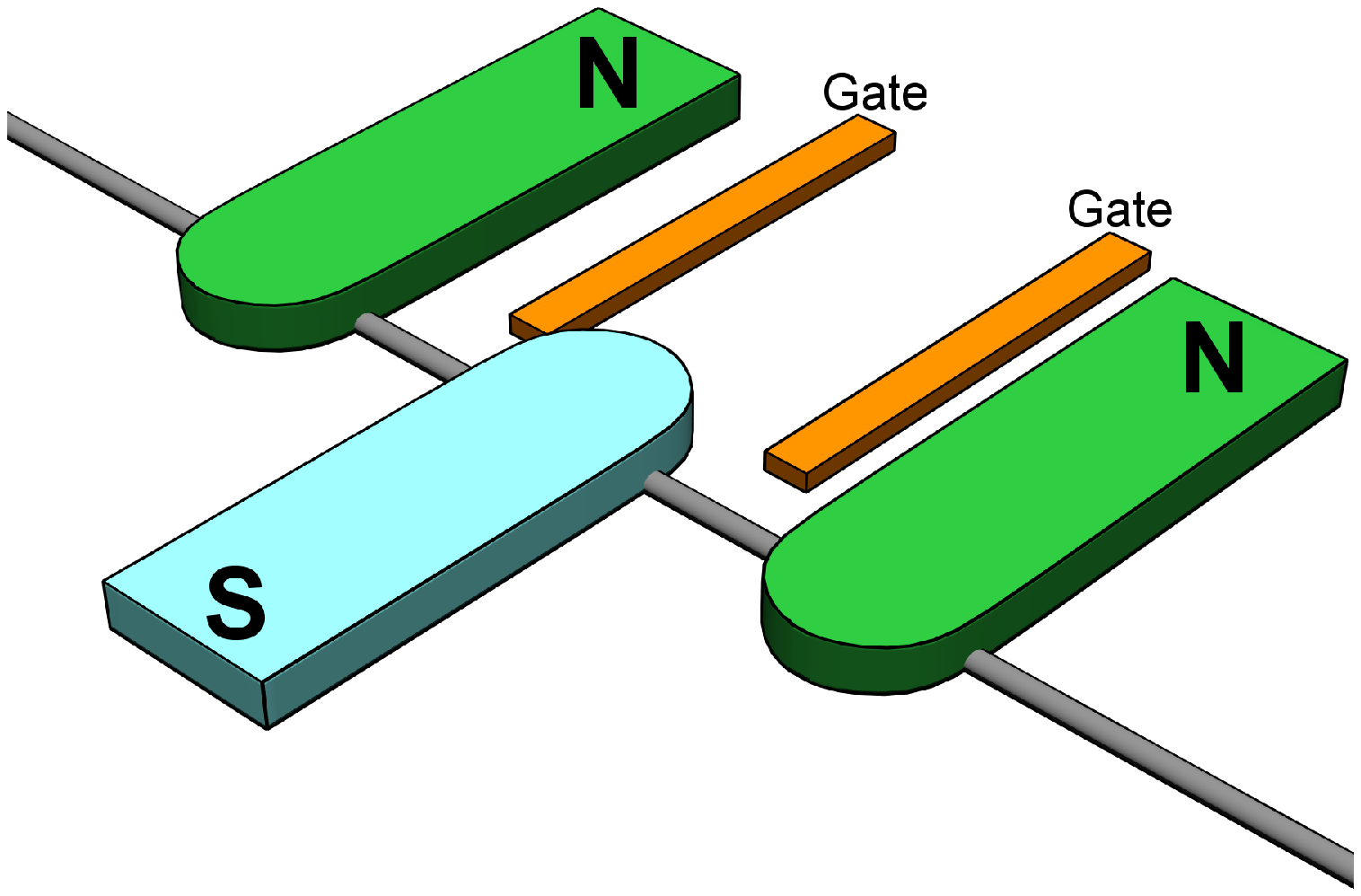} & (b)& \includegraphics[width=3.6cm]{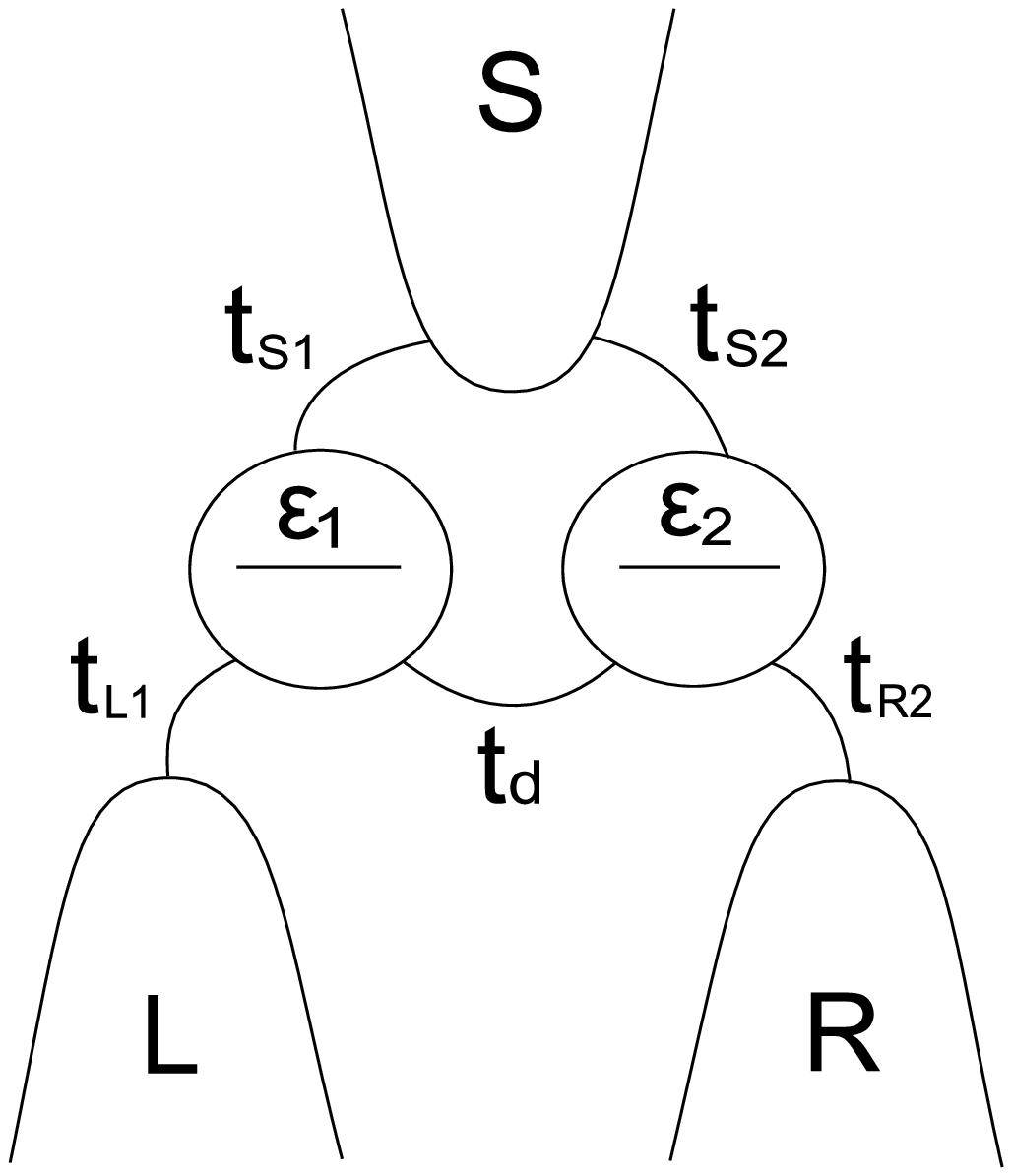} 
 \end{tabular}}
\caption{
Cooper-pair beam-splitter setup: (a) representation of the experimental device where a nanowire/nanotube is coupled to normal and superconducting leads, and (b) schematic representation of the double-dot model considered in the text (with the corresponding notations).}
\label{fig-setup}
\end{figure}

Let us now illustrate this procedure on the particular example of the double-dot Cooper-pair beam splitter, whose setup is recalled in Fig.~\ref{fig-setup}. 

\subsection{Derivation of the self-energy and vertex function}

Our system now consists of a superconducting lead at potential $V_S=0$, coupled to two quantum dots (labeled 1 and 2), which are each connected to a different normal electrode (at potential $V_L$ and $V_R$ respectively). We will be interested in the branching currents $I_{L1}$ and $I_{R2}$ as well as their cross-correlations. Here $I_{L1}$ corresponds to the current between dot 1 and the left normal lead, while $I_{R2}$ flows between dot 2 and the right normal lead.

In what follows we focus on the first term in the diagrammatic expansion of the Luttinger-Ward functional so that our conserving approximation amounts to writing
\begin{align}
\Phi \left[ \cG \right] = \frac{i}{2} \int dt \sum_{\substack{\sigma, \sigma' \\ \eta, \alpha}} \tau_z^{\eta \eta} U_{\alpha}  \sigma_z^{\sigma\sigma} \sigma_z^{\sigma' \sigma'}   \cG^{\eta \eta}_{\substack{\alpha \alpha \\ \sigma \sigma'}} (t,t) \cG^{\eta \eta}_{\substack{\alpha \alpha \\ \bar{\sigma} \bar{\sigma}'}} (t,t) ,
\label{eq-Phi1storder}
\end{align}
where $\bar{\sigma} = - \sigma$. 

Following the procedure outlined in the previous section, we first obtain the interacting self-energy, which in frequency space writes
\begin{equation}
\cSigma_{C,\mu_1 \mu_2} (\omega) = i U_{\alpha_1} \delta_{\alpha_1 \alpha_2} \tau_z^{\eta_1 \eta_2} \sigma_z^{\sigma_1 \sigma_1} \sigma_z^{\sigma_2 \sigma_2} \int \frac{d \Omega}{2\pi} \cG_{\mu_1 \mu_2}  (\Omega) .
\end{equation}
One readily sees from this expression that at this level of approximation the interacting self-energy in Fourier space is frequency independent, and can thus be viewed as a renormalization of the dots energies. However, this constant still needs to be evaluated self-consistently, through the Dyson equation~\eqref{eq-Dyson}, as it depends on the full single-particle Green's function $\cG$.

The irreducible vertex is further obtained following Eq.~\eqref{eq-irred}, and takes the following simple frequency independent form in Fourier space
\begin{align}
\bar{\Lambda}_{\mu_1 \mu_2 \mu_3 \mu_4} =& i \tau_z^{\eta_1 \eta_2} \delta_{\eta_1 \eta_3} \delta_{\eta_1 \eta_4} U_{\alpha_1} \delta_{\alpha_1 \alpha_2} \delta_{\alpha_1 \alpha_3} \delta_{\alpha_1 \alpha_4} \nonumber \\
& \times \sigma_z^{\sigma_1 \sigma_1} \sigma_z^{\sigma_3 \sigma_3} \delta_{\sigma_1 \bar{\sigma}_4}   \delta_{\sigma_2 \bar{\sigma}_3} ,
\label{eq-irrvertex}
\end{align}
where we used the energy conservation to introduce $\cLambda (\omega_1,\omega_2,\omega_3,\omega_4) = 2 \pi \delta \left( \omega_1-\omega_2-\omega_3+\omega_4 \right) \bar{\Lambda}$. The result of Eq.~\eqref{eq-irrvertex} is nothing but the bare interaction vertex where one recognizes the properties of locality in time (all $t_i$ and $\eta_i$ are equal) and space (all $\alpha_i$ are equal), as well as the spin conservation (the sum of incoming and outgoing spin projections is zero).

\begin{figure*}[tb]
\centering
 \resizebox{.94\textwidth}{!}{\includegraphics*{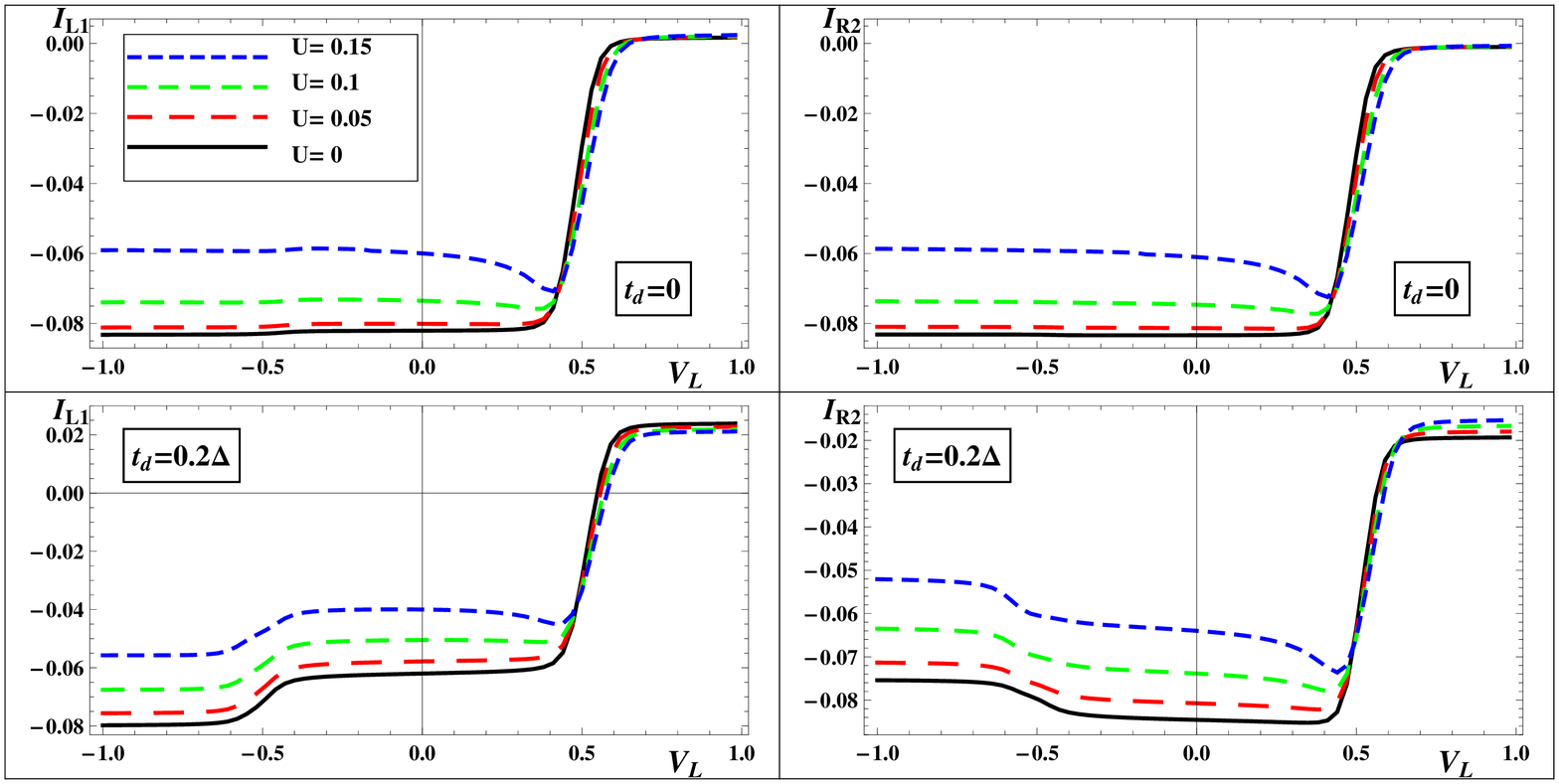}}
\caption{
Currents $I_{L1}$ and $I_{R2}$ (in units of $I_0 = e \Delta/\hbar$) as a function of $V_L$, for different values of the interaction $U=0, 0.05, 0.1, 0.15 \Delta$, and the parameters (in units of $\Delta$) $\beta=100$, $\epsilon_1=0.5$, $\epsilon_2=-0.5$, $V_R=-0.7$ and $t_{L1} = t_{S1} = t_{S2} = t_{R2} = 0.2$. The upper panels correspond to $t_d=0$, while the lower panels are computed for $t_d=0.2 \Delta$.}
\label{fig-currents}
\end{figure*} 

\begin{widetext}
Substituting this expression for the irreducible vertex back into the Bethe-Salpeter equation \eqref{eq-BetheSalpeter} and moving to Fourier space, one notices that the frequency-dependent full vertex function depends on a single frequency argument, so that by further introducing $\cGamma (\omega_1,\omega_2,\omega_3,\omega_4) = 2 \pi \delta \left( \omega_1-\omega_2-\omega_3+\omega_4 \right) \bar{\Gamma} \left(\omega_2-\omega_1 \right)$, we are left with
\begin{align}
 \bar{\Gamma}_{\mu_1\mu_2\mu_3\mu_4} \left( \omega_2 - \omega_1 \right) = \bar{\Lambda}_{\mu_1 \mu_2 \mu_3 \mu_4} + \int \frac{d\Omega}{2\pi} \sum_{\mu_5 \ldots \mu_8} \bar{\Lambda}_{\mu_1\mu_2\mu_5\mu_6} \cG_{\mu_5 \mu_7} (\Omega) \cG_{\mu_8 \mu_6} (\Omega+\omega_2-\omega_1) \bar{\Gamma}_{\mu_7\mu_8\mu_3\mu_4} \left( \omega_2-\omega_1\right) .
 \label{eq-fullvertex}
\end{align}
\end{widetext}

In order to extract an explicit expression for the full vertex, one needs to invert Eq.~\eqref{eq-fullvertex}, which cannot be straightforwardly achieved since we are dealing with $4^{\text{th}}$ order tensors. To circumvent this issue, we propose to construct a matrix representation of the full and irreducible vertices in an enlarged NDK$\otimes$NDK space, by combining two by two the four labels of these tensor elements.\footnote{There are obviously three choices for such a construction. This choice is ultimately guided by the irreducibility channel considered for the Bethe-Salpeter equation.}
Indeed, the matrix representation of the single-particle Green's function relies on identifying a single coordinate in NDK space $n_i$ (running from 1 through 8) out of the values of the Nambu, dot and Keldysh components $\left\{\sigma_i,\alpha_i,\eta_i\right\}$. In the same spirit, we construct enlarged $8^2 \times 8^2$ matrices to represent the full and irreducible vertices, in which the coordinates are obtained from a linear combination of the $n_i$ according to
\begin{equation}
\cG^{\eta_1 \eta_2}_{\substack{\alpha_1 \alpha_2 \\ \sigma_1 \sigma_2}} = \left[\cG\right]_{n_1 n_2} 
~ \Rightarrow ~
\bar{\Gamma}^{\eta_1 \eta_2 \eta_3 \eta_4}_{\begin{subarray}{1}
\alpha_1 \alpha_2 \alpha_3 \alpha_4\\ \sigma_1 \sigma_2 \sigma_3 \sigma_4
\end{subarray}} = \bar{\Gamma}_{n_1 n_2 n_3 n_4} = \left[\bar{\Gamma}\right]_{N_1 N_2}
\end{equation}
with $N_1=8 (n_1-1)+n_2$ and $N_2=8 (n_3-1)+n_4$. 
The resulting $64\times64$ matrices, which we hereby denote with brackets, ultimately lead to the following explicit form for the full-vertex function 
\begin{align}
\left[\bar{\Gamma} (\omega_2 - \omega_1)\right]  & = \left[\bar{\Lambda}\right] \left\{ \left[ \mathds{1} \right] - \left[\Pi(\omega_2-\omega_1)\right] \times \left[\bar{\Lambda}\right] \right\}^{-1} ,
\end{align}
where we introduced the generalized polarization bubble $\left[\Pi \right]$ as
\begin{equation}
\left[\Pi (\omega)\right] = \int \frac{d\Omega}{2\pi} \left[ \check{G} (\Omega) \otimes \check{G}^T (\Omega+\omega) \right] ,
\end{equation}
and $\check{G}^T$ is the transpose of the matrix $\check{G}$ in NDK space.

\subsection{Numerical results}

We computed numerically the currents and crossed correlations as a function of the left lead voltage $V_L$ (with fixed $V_R$), for different values of the interaction parameter $U_1 = U_2 = U$. We focused on the most favorable regime for CAR, i.e. when the dots energies are chosen antisymmetric with respect to the superconductor chemical potential, and considered the situations of a negligible and a strong direct tunneling. 

We first solve self-consistently the Dyson equation in order to compute the single-particle Green's function $\cG$ and derive the currents $I_{L1}$ and $I_{R2}$ from Eq.~\eqref{eq-Current}, which we present in Fig.~\ref{fig-currents}. In absence of direct tunneling, the two currents are almost identical, while for a finite $t_d$, the current difference is significant, due to the opening of a new conduction channel between the normal leads.

As a function of $U$, one readily sees that the amplitude of all currents gets reduced, already for small values of the interaction strength. This reduction does not equally affect the whole range of voltages. In particular the region of high voltages, where the current is dominated by electron co-tunneling is only marginally modified. However, the current reduction in the presence of interactions is sizable in the low-voltage CAR-dominated regime.  Moreover, the effects are qualitatively similar whether a direct tunneling is present or not. These are purely static single-particle effects, and can be understood as a detuning of the resonances in the dots density of states,\cite{non-int} a direct consequence of the constant interacting self-energy. Note that while the latter is frequency-independent, it does depend on $V_L$ (through self-consistency), explaining why different regions of the curves are not equally affected.

The most interesting and non-trivial results come from the calculation of the current cross-correlations as both one- and two-particle effects are present in this case. In what follows, we focus on the zero-frequency current-current correlation $\mathcal{S}_{i\alpha,j\beta} (0)$ defined as
\begin{equation}
\mathcal{S}_{i\alpha,j\beta} (0) = \int dt \left( \langle I_{i\alpha}^- (t) I_{j\beta}^+ (0) \rangle - \langle I_{i\alpha} (t) \rangle \langle  I_{j\beta} (0) \rangle \right) .
\end{equation}

\begin{widetext}
Substituting the expressions \eqref{eq-Current} for the branching currents and \eqref{eq-CurrCorr} for the current-current correlators, and using Eq.~\eqref{eq-GammatoK} to replace the two-particle Green's function in terms of the full vertex, the zero-frequency correlations become
\begin{multline}
\mathcal{S}_{i\alpha,j\beta} (0) = e^2 \sum_{\sigma \sigma'} \sigma_z^{\sigma \sigma} \sigma_z^{\sigma' \sigma'} \sum_{\mu_1 \mu_2} 
 \left\{ \int \frac{d\omega}{2\pi} 
\left[\cSigma_{i,\nu \mu_1} \cSigma_{j,\nu' \mu_2} \cG_{\mu_1 \nu'} \cG_{\mu_2 \nu}  
 -\cSigma_{i,\nu \mu_1} \cSigma_{j,\mu_2 \nu'} \cG_{\mu_1 \mu_2} \cG_{\nu' \nu}   \right. \right. \\
\left. -\cSigma_{i,\mu_1 \nu} \cSigma_{j,\nu' \mu_2} \cG_{\nu \nu'} \cG_{\mu_2 \mu_1}   
  +\cSigma_{i,\mu_1 \nu} \cSigma_{j,\mu_2 \nu'} \cG_{\nu \mu_2} \cG_{\nu' \mu_1}  \right] (\omega) \\
+ \int \frac{d\omega_1}{2\pi}  \frac{d\omega_2}{2\pi} \sum_{\mu_5,\mu_6,\mu_7,\mu_8} \bar{\Gamma}_{\mu_5 \mu_6 \mu_7 \mu_8} (\omega_2-\omega_1) 
  \left[ \left( \cSigma_{i,\nu \mu_1} \cG_{\mu_1 \mu_5} \cG_{\mu_7 \nu} \right) (\omega_1) 
  \left(\cSigma_{j,\nu' \mu_2} \cG_{\mu_2 \mu_8} \cG_{\mu_6 \nu'} \right) (\omega_2)
 \right. \\
 - \left(\cSigma_{i,\nu \mu_1} \cG_{\mu_1 \mu_5} \cG_{\mu_7 \nu} \right) (\omega_1) 
 \left(\cSigma_{j,\mu_2 \nu'} \cG_{\nu' \mu_8} \cG_{\mu_6 \mu_2} \right) (\omega_2) 
 - \left(\cSigma_{i,\mu_1 \nu} \cG_{\nu \mu_5} \cG_{\mu_7 \mu_1} \right) (\omega_1) 
 \left(\cSigma_{j,\nu' \mu_2} \cG_{\mu_2 \mu_8} \cG_{\mu_6 \nu'} \right)    (\omega_2) 
 \\
 \left. + \left(\cSigma_{i,\mu_1 \nu} \cG_{\nu \mu_5} \cG_{\mu_7 \mu_1} \right) (\omega_1) 
 \left(\cSigma_{j,\mu_2 \nu'} \cG_{\nu' \mu_8} \cG_{\mu_6 \mu_2} \right) (\omega_2)  \right] \bigg\} .
\end{multline} 
\end{widetext}

Using the matrix representation introduced above, the cross-correlations can be recast in a much more manageable form as
\begin{align}
\mathcal{S}_{L1,R2}(0) &= e^2 \int \frac{d\omega}{2\pi} \trace{NDK}  \left[  \cXi_{L1}^- \cG \cXi_{R2}^+ \cG \right] (\omega)  \nonumber \\
&+ e^2 \int \frac{d\omega_1}{2\pi} \frac{d\omega_2}{2\pi} \mathrm{Tr}_{\mathrm{NDK}^2} \Big\{ \left[\bar{\Gamma} (\omega_2-\omega_1) \right]  \nonumber \\
& \times \left.  \left[\left(\cG \cXi_{L1}^- \cG\right) (\omega_1)  \otimes \left( \cG \cXi_{R2}^+ \cG\right)^T (\omega_2)  \right] \right\} ,
\label{eq-SL1R2}
\end{align}
where we introduced the trace $\mathrm{Tr}_{\mathrm{NDK}^2}$ over our enlarged NDK$\otimes$NDK matrices and defined the compact notations
\begin{align}
\cXi_{i\alpha}^\eta (\omega) &= \left(\pi_{\alpha \eta} \cSigma_i (\omega) -  \cSigma_i (\omega) \pi_{\alpha \eta} \right) \\
\pi_{\alpha \pm} &= \frac{1}{2} \left(\mathds{1} \pm \tau_z\right) \otimes {\cal I}_\alpha \otimes \sigma_z .
\end{align}
The first term on the r.h.s of Eq.~\eqref{eq-SL1R2} is similar to the expression obtained in the absence of interactions,\cite{non-int} only here it involves the full dots electrons single-particle Green's function. The second term however corresponds to truly two-particle effects.

\begin{figure}[b]
\centering
 \resizebox{.48\textwidth}{!}{\includegraphics*{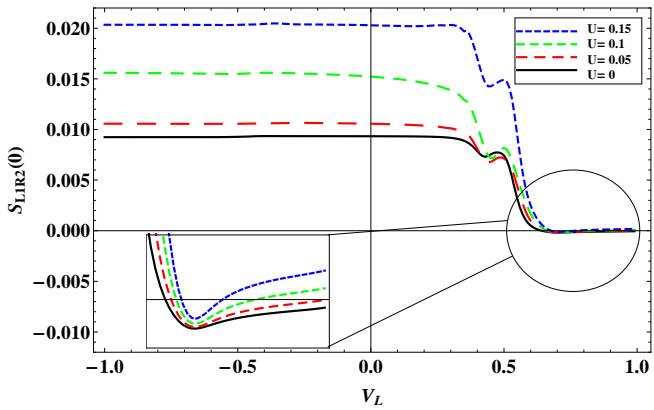}}
\caption{
Zero-frequency current cross-correlations (in units of $S_0 = e^2 \Delta/\hbar$) as a function of $V_L$, for $U=0, 0.05, 0.1, 0.15 \Delta$, and the same parameters as in Fig.~\ref{fig-currents}, in the absence of direct tunneling.}
\label{fig-crossnotd}
\end{figure} 

In Fig.~\ref{fig-crossnotd}, we show the results obtained for the zero-frequency current-current correlations in the absence of direct tunneling. Like in the non-interacting case, one can typically isolate two relevant regimes of voltages. At low voltage ($V_L < \epsilon_1$), transport is dominated by CAR processes and we observe positive cross-correlations, which are approximately constant. Increasing the interaction parameter leads to strongly enhanced correlations in this region. This enhancement cannot be solely attributed to the renormalization of $\cG$ due to interactions. We could check that the vertex-dependent term in Eq.~\eqref{eq-SL1R2} contributes substantially to the final result, therefore suggesting that interaction-induced two-particle effects are particularly important in the Cooper-pair beam-splitter setup.

\begin{figure}[b]
\centering
 \resizebox{.48\textwidth}{!}{\includegraphics*{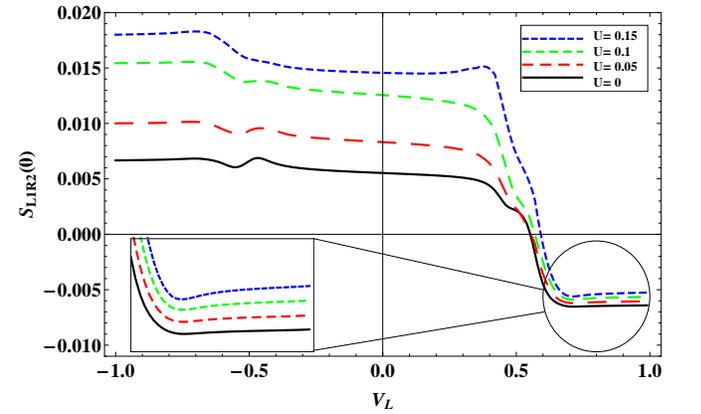}}
\caption{
Zero-frequency current cross-correlations (in units of $S_0 = e^2 \Delta/\hbar$) as a function of $V_L$, for $U=0, 0.05, 0.1, 0.15 \Delta$, and the same parameters as in Fig.~\ref{fig-currents}, in the presence of direct tunneling, $t_d=0.2 \Delta$.}
\label{fig-crossandtd}
\end{figure}

In the region of high voltage ($V_L > \epsilon_1$), we came across a somewhat counter-intuitive result. In this regime, CAR is greatly unfavored and one expects electron co-tunneling to be the dominant process, as confirmed by the current plots of Fig.~\ref{fig-currents} where $I_{L1}$ and $I_{R2}$ are opposite in this region therefore signaling the transfer of electrons from the left to the right normal lead. One would thus expect to observe negative correlations, as was the case in the absence of interactions. However, paying close attention to the high-voltage region (see the magnified area of Fig.~\ref{fig-crossnotd}), we see that the effect of interactions, although weak, is sufficient to flip the overall sign, resulting in slightly positive correlations.

When a direct tunneling is allowed between dots, there is still an important enhancement of positive correlations at low voltage, even for weak interaction strength (see Fig.~\ref{fig-crossandtd}). Also, in contrast with the constant behavior observed for $t_d=0$, a more pronounced feature develops around $V_L \simeq - \epsilon_1$. This is reminiscent of what was observed in the non-interacting case, and can be attributed to density of states effects\cite{non-int}. For voltages $V_L > \epsilon_1$, interactions have a weaker effect, and tend to reduce the amplitude of cross-correlations, without causing any change of sign.

Comparing the results of Figs.~\ref{fig-crossnotd} and \ref{fig-crossandtd}, it becomes clear that while a finite $t_d$ tends to shift the correlations towards negative values, a finite $U$ have the opposite effect of shifting them towards positive ones, thus revealing the competition between interactions and direct tunneling.

\section{Conclusion and outlook} \label{sec:conclusion}

In conclusion, we presented a conserving approach to study currents and their correlations in interacting out-of-equilibrium mesoscopic systems. This Kadanoff-Baym-Keldysh formalism amounts to computing the single- and two-particle dots electrons Green's functions from a given truncation of the Luttinger-Ward functional. These are then used to express the branching currents (in the same spirit as the Meir-Wingreen formula\cite{meir_wingreen}) as well as their correlations, therefore extending the Fisher-Lee/Landauer-B\"uttiker formula\cite{fisher_lee} for the noise to the case of an interacting system. Applying this formalism to the double-dot Cooper-pair beam splitter setup, we could study the effect of Coulomb interaction on the currents and cross-correlations, showing that even weak interactions have an important effect in the CAR-dominated regime, in particular favoring positive cross-correlations in the antisymmetric case. Our results also exhibit the competition between local Coulomb interaction and direct inter-dot tunneling, which affect the current correlations in an opposite manner.

The present work can be extended in various ways. For example, the formalism can also treat a non-local Coulomb interaction between dots. This would yield new terms in $\Phi\left[\cG\right]$, and therefore in the self-energy and irreducible vertex, which can be easily accounted for. But the most natural extension consists in including a second set of diagrams into the Luttinger-Ward functional (see Fig.~\ref{fig-diagrams}a). In particular, this second order contribution would lead to a newly acquired frequency dependence of the interacting self-energy and irreducible vertex, and thus to new features associated with these retardation effects. However, this would also make the Bethe-Salpeter equation no longer invertible analytically, therefore requiring a self-consistent numerical solution.


\begin{thebibliography}{99}

\bibitem{hofstetter} L. Hofstetter, S. Csonka, J. Nyg\aa rd and C. Sch\"onenberger, Nature (London) {\bf 461}, 960 (2009); L. Hofstetter et al., Phys. Rev. Lett. {\bf 107}, 136801 (2011).

\bibitem{herrmann} L. G. Herrmann, F. Portier, P. Roche, A. L. Yeyati, T. Kontos and C. Strunk, Phys. Rev. Lett. {\bf 104}, 026801 (2010). 

\bibitem{anatram_datta} M.P. Anantram and S. Datta, Phys. Rev. B {\bf 53}, 16390 (1996). 

\bibitem{lesovik_martin_blatter} G. B. Lesovik, T. Martin, and G. Blatter, Eur. Phys. J. B {\bf 24}, 287 (2001).

\bibitem{bouchiat} V. Bouchiat, N. M. Chtchelkatchev, D. Feinberg, G. B. Lesovik, T. Martin, and J. Torr\`es, Nanotechnology {\bf 14}, 77 (2003). 

\bibitem{torres-martin} J. Torr\`es and T. Martin, Eur. Phys. J. B {\bf 12}, 319 (1999).

\bibitem{deutscher_feinberg} G. Deutscher and D. Feinberg, Appl. Phys. Lett. {\bf 76}, 487 (2000).

\bibitem{falci_feinberg_hekking} D. F. G. Falci, D. Feinberg, and F. W. J. Hekking, Europhys. Lett. {\bf 54}, 255 (2001).

\bibitem{beckman_prl} D. Beckmann, H. B. Weber, and H. v. L\"ohneysen, Phys. Rev. Lett. {\bf 93}, 197003 (2004).

\bibitem{klapwijk} S. Russo, M. Kroug, T. M. Klapwijk, and A. F. Morpurgo, Phys. Rev. Lett. {\bf 95}, 027002 (2005).

\bibitem{chandrasekhar} P. Cadden-Zimansky and V. Chandrasekhar, Phys. Rev. Lett. {\bf 97}, 237003 (2006).

\bibitem{sukhorukov_recher_loss} P. Recher, E. V. Sukhorukov, and D. Loss, Phys. Rev. B {\bf 63}, 165314 (2001).

\bibitem{borlin} J. B\"orlin, W. Belzig and C. Bruder, Phys. Rev. Lett. {\bf 88}, 197001  (2002).

\bibitem{yeyati_bergeret} A. Levy Yeyati, F. S. Bergeret, A. Martin-Rodero, and T. Klapwijk, Nat. Phys. {\bf 3}, 455 (2007).

\bibitem{zaikin_golubev} D. S. Golubev and A. D. Zaikin, Phys. Rev. B {\bf 82}, 134508 (2010).  

\bibitem{eldridge} J. Eldridge, M. G. Pala, M. Governale and J. K\"onig, Phys. Rev. B {\bf 82}, 184507 (2010).

\bibitem{hershfield} S. Hershfield, Phys. Rev. B {\bf 46}, 7061 (1992).

\bibitem{lopez} R. L\'opez, R. Aguado and G. Platero, Phys. Rev. B {\bf 69}, 235305 (2004).

\bibitem{thielmann} A. Thielmann, M. H. Hettler, J. K\"onig, and G. Sch\"on, Phys. Rev. Lett. {\bf 95}, 146806 (2005).

\bibitem{fujii} T. Fujii, J. Phys. Soc. Jpn. {\bf 79}, 044714 (2010).

\bibitem{simon} C.P. Moca, P. Simon, C.H. Chung, and G. Zar\'and, Phys. Rev. B {\bf 83}, 201303(R) (2011).

\bibitem{baym-kadanoff} L.P. Kadanoff and G. Baym, {\it Quantum Statistical Mechanics} (Benjamin, New York, 1962).

\bibitem{non-int} D. Chevallier, J. Rech, T. Jonckheere and T. Martin, Phys. Rev. B {\bf 83}, 125421 (2011).

\bibitem{luttinger-ward} J. M. Luttinger and J. C. Ward, Phys. Rev. {\bf 118}, 1417 (1960).

\bibitem{meir_wingreen} Y. Meir and N. S. Wingreen, Phys. Rev. Lett. {\bf 68}, 2512 (1992).

\bibitem{fisher_lee} R. Landauer, IBM J. Res. Dev. {\bf 1}, 223 (1957); D. S. Fisher and P. A. Lee, Phys. Rev. B {\bf 23}, 6851 (1981); M. B\"uttiker, Phys. Rev. B {\bf 46}, 12485 (1992).


\end{thebibliography}
\end{document}